\documentclass[12pt]{article}
\begin{document}
\begin{center}
{\large {SOLDERING FORMALISM IN NONCOMMUTATIVE FIELD THEORY:\\
A BRIEF NOTE}}
\vskip 2cm
Subir Ghosh\\
\vskip 1cm
Physics and Applied Mathematics Unit,\\
Indian Statistical Institute,\\
203 B. T. Road, Calcutta 700108, \\
India.

\vskip 3cm
 Abstract:\\
\end{center}
In this paper, I develop the Soldering formalism in a new domain -
the noncommutative planar field theories. The Soldering mechanism
fuses two distinct theories showing opposite or complimentary
properties of some symmetry, taking into account the interference
effects. The above mentioned symmetry is hidden in the composite
(or soldered) theory. In the present work it is shown that a pair
of noncommutative Maxwell-Chern-Simons theories, having opposite
signs in their respective topological terms, can be consistently
soldered to yield the Proca model (Maxwell theory with a mass
term) with corrections that are at least quadratic in the
noncommutativity parameter. We further argue that this model can
be thought of as the noncommutative generalization of the Proca
theory  of  ordinary spacetime. It is well-known that abelian
noncommutative gauge theory bears a close structural similarity
with non-abelian gauge theory. This fact is manifested in a
non-trivial way if the present work is compared with existing
literature, where soldering of non-abelian models are discussed.
Thus the present work further establishes the robustness of the
soldering programme. The subtle role played by gauge invariance,
(or the lack of it), in the above soldering process, is revealed
in an interesting way.


\newpage
In recent years, Non-Commutative (NC) field theories \cite{sw,rev}
and in particular NC gauge theories  have generated a lot of
interest due to their appearance in the low energy limit in a
system of open strings ending on $D$-branes, in the presence of a
background field. The $D$-branes inherit the noncommutativity in
the string boundaries. Thus the field theories living on the
$D$-branes can be described by NC field theories, which can yield
string theoretic results in certain limits. On the other hand, by
itself NC field theory is a fascinating subject. Even though in NC
quantum field theory, the basic computational scheme remains
essentially  the same as that of quantum field theory in ordinary
spacetime, qualitatively distinct behavior is observed in the
former.  Some of the novel features of NC quantum field theories
are UV/IR mixing \cite{min1} induced by non-locality, presence of
solitons \cite{min2} in higher dimensional scalar theories, dipole
like elementary excitations \cite{jab}, etc. This motivates
further study of different aspects of field theories in the
context of NC spacetime. In the present Note, we will concentrate
on some specific models in 2+1-dimensional NC spacetime.

Investigations in the context of quantum field theories living in
2+1-dimensional ordinary (commutative) spacetime have proved to be
rewarding in the past \cite{djt,fs,sg}. There exist several
physically relevant models, such as Maxwell-Chern-Simons (MCS)
model, Self-Dual (SD) models, fermions in interaction with gauge
fields (massive Thirring model), that are closely interrelated and
enjoy non-trivial duality (or equivalence) relations between
operators of the respective theories. The MCS and SD models, along
with their dual nature, have been studied exhaustively in
\cite{djt}. Their connection with the fermion theories via
bosonization (in the large fermion mass limit) was elucidated in
\cite{fs}. A unified analysis of all these models can be found in
\cite{sg}.

A number of works, spanning all the above topics, pertaining to NC
generalization, have appeared recently \cite{sg1,sg2,of,bot}. An
NC generalization of the MCS model, obtained by exploiting the
inverse Seiberg-Witten map \cite{sw} and its subsequent duality
with the NC SD model was shown by the author in \cite{sg1}. In the
above mentioned NC SD model, the Chern-Simons term is structurally
identical to the Chern-Simons
 term in ordinary spacetime. Bosonization of the NC massive
Thirring model, in the large fermion mass limit, was carried
through in \cite{sg2}, which reproduced a variant of the NC SD
model, where the Chern-Simons term is the NC Chern-Simons term. Hence it was concluded
(see Ghosh in \cite{sg1}) that duality between the massive
Thirring model and MCS model (in large fermion mass limit), a
property valid in ordinary spacetime \cite{djt,fs}, is lost in NC
spacetime. However, later it was shown in \cite{bot} that the
above chain of duality can be maintained in NC spacetime as well,
provided one considers an alternate version of the NC MCS model,
proposed in \cite{of}, consisting of NC Maxwell and NC
Chern-Simons terms. Interestingly, \cite{bot} shows that this NC
MCS model is actually dual to the model obtained in \cite{sg2} via
bosonization, thereby completing the chain of dualities. All the
above results are valid for the lowest non-trivial order in
$\theta_{\mu\nu}$- the noncommutativity parameter.

As far as $O(\theta)$ computations are concerned, both  the
above definitions of NC MCS theory in \cite{sg1} and \cite{of} are
equally viable but distinct alternatives, with \cite{of} probably
being the more popular one. On the other hand, the issue of NC
extension of the SD model is tricky. If we follow the approach of
the Seiberg-Witten map \cite{sw}, our way of defining the NC SD
model in \cite{sg1}  appears to be the natural one since in the
absence of any (manifest) gauge invariance, the Seiberg-Witten map
should not come in to the picture. Since the SD model in ordinary
spacetime is a quadratic theory with no gauge symmetry, there will
be no significant effects of noncommutativity. This is because the
$\theta$-dependent term will come only from the $*$-product of two
operators. These contributions are ignored assuming that total
derivative terms can be dropped from the action. Hence the model
in question in \cite{sg2,bot}, consisting of a mass term and NC
Chern-Simons term, should not be thought of as NC SD model. In the
present paper, our reasoning will be corroborated further in the
context of another model of a similar nature - the Proca model.

It is now time to put our work in its proper perspective. The
Soldering formalism \cite{st,bw,sg3}- to be explained below - has
been used extensively, (see \cite{w} for un updated review and
references therein), in the context of theories in ordinary
spacetime. We demonstrate that it is adaptable to noncommutative
spacetime as well, leading to interesting and non-trivial results.
The present work is concerned with $O(\theta )$ modifications
only. The noncommutativity brings in additional features which can
be directly related to similar behavior in soldering in
non-abelian gauge theories in ordinary spacetime \cite{iw}.

Furthermore, the noncommutative soldering, in the present case,
generates a particularly simple model, which we would like to
interpret as the Proca model in NC field theory framework.
  We will argue that the NC version of the Proca model
should contain the mass term and  {\it{ordinary}} Maxwell term,
(with possible $O(\theta^{2})$ corrections, $\theta_{\mu\nu}$
being the noncommutativity parameter ), and not the NC Maxwell
term, for reasons exactly similar as above. In the NC theory
considered here, the NC Proca model appears in an exactly analogous
situation where the Proca model emerges in ordinary spacetime. Our
framework in demonstrating the above will be the Soldering
formalism .

Let us briefly introduce the idea of Soldering formalism
\cite{st}. Combination of two distinct models to construct a
single composite model is interesting, especially if both the
initial and final theories are physically relevant. The above
result clearly shows a direct connection between the parent and
daughter theories. However, pursuing this idea in a generic case,
in a  systematic way, is indeed nontrivial. The Soldering
formalism precisely does this job for a particular class of a pair
of models, which manifest the {\it{dual}} aspects of some
symmetry, such as chirality, self-duality \cite{bw}, etc.. The
soldering procedure can be carried through for such a particular
pair in a well defined manner and the resulting soldered model, in
a certain sense, hides the above mentioned symmetry. In fact, the
above mentioned equivalence between the parent and daughter
theories is quite deep rooted. This can be established \cite{sg3}
in an alternative Canonical Transformation  scheme, whereby in a
Hamiltonian framework, the soldered model can be broken up into
the parent dual models. Hence, the Soldering formalism and the
Canonical Transformation prescription are complimentary to each
other.

I start by introducing the SD-MCS duality in conventional
space-time. For convenience we follow the notations  and metric
$(g^{\mu\nu}=diag~(1,-1,-1))$ of \cite{djt}. The self-dual (or
anti self-dual) lagrangians, consisting of the ordinary and
topological mass terms are,
\begin{equation}
{\cal {L}^{\pm}_{SD}}=\frac{1}{2}f^\mu f_\mu\pm\frac{1}{2M}\epsilon ^{\alpha\beta\gamma}f_\alpha\partial _\beta f_\gamma .
\label{sd}
\end{equation}
These two models can be generated via bosonization \cite{fs},
(in the large fermion mass limit), from two distinct fermion
theories of mass $M$, having opposite chiralities. On the other
hand, the (corresponding dual) MCS models \cite{djt} are described
by,
\begin{equation}
{\cal {L}^{\pm}_{MCS}}=-\frac{1}{4}F^{\alpha\beta}F_{\alpha\beta}\pm \frac{M}{2}\epsilon^{\alpha\beta\gamma}A_{\alpha}\partial_{\beta}A_{\gamma};~~
F_{\alpha\beta}=\partial_{\alpha}A_{\beta}-\partial_{\beta}A_{\alpha}.
\label{mcs1}
\end{equation}
Note that total derivative terms in the lagrangian will be dropped
throughout the present (classical) discussion. Clearly
(\ref{mcs1}) is invariant under the gauge transformation,
\begin{equation}
A_{\mu}\rightarrow A_{\mu}+\partial_{\mu}\lambda ,
\label{tr1}
\end{equation}
whereas no such manifest symmetry exists for (\ref{sd}). However,
one can solve the equations of motion and constraints for both of
the above models and show that there exists the identification
\cite{djt},
$$f_{\mu}=\epsilon_{\mu\nu\tau}\partial^{\nu}A^{\tau},$$
that reduces one model to the other.

Before discussing the role of Soldering in the present context, let me introduce
the Soldering mechanism \cite{bw} in an explicit way for a generic situation. Here an
iterative Noether procedure is exploited to lift the global
symmetries  of the constituent models to a  local symmetry in the
composite model. The general prescription is to express the
variations of the actions of the models, ${\cal {A}}^{+}(A)$ and
${\cal {A}}^{-}(B)$, (that are to be soldered), in the following
form,
\begin{equation}
\delta {\cal {A}}^{+}(A)=\int d^{3}x J^{+}_{\mu}(A)\partial^{\mu}a,~~
\delta {\cal {A}}^{-}(B)=\int d^{3}x J^{-}_{\mu}(B)\partial^{\mu}a,
\label{s1}
\end{equation}
where $J^{+}_{\mu}(A)$ and $J^{-}_{\mu}(B)$ denote the generic
Noether currents corresponding to the global invariances under,
\begin{equation}
\delta A=\delta B = a.
\label{s2}
\end{equation}
Next one introduces an auxiliary field $C$, with a particular local transformation
\begin{equation}
\delta C\approx f(a).
\label{s3}
\end{equation}
 such that the following action
\begin{equation}
{\cal {A}}(A,B,C)={\cal {A}}^{+}(A)+{\cal {A}}^{-}(B)+{\cal{W}}(A,B,C),
\label{ss3}
\end{equation}
is invariant under the local transformations given in
(\ref{s2},\ref{s3}). The last term in the action,
${\cal{W}}(A,B,C)$, incorporates the interference effects. It is
of such a form that the variational equation of motion for $C$ is
an algebraic one for $C$. Thus $C$ can be eliminated classically
from (\ref{s3}), thereby yielding the cherished action ${\cal
{A}}^{S}(G)$ for the soldered model, where the fields $A$ and $B$
occur in a gauge invariant combination $G\equiv A-B,~\delta G=0$.

In ordinary spacetime, application of the soldering mechanism for
the self and anti self-dual models, (or there MCS versions),
induces the Proca model \cite{bw},
\begin{equation}
L(G)=-\frac{1}{4}F^{\mu\nu}(G)F_{\mu\nu}(G)+\frac{m^{2}}{2}G^{\mu}G_{\mu};~~G_{\mu}=\frac{1}{\sqrt{2}}(A_{\mu}-B_{\mu}).
\label{s4}
\end{equation}
As mentioned before, we will concentrate on the analogous
phenomenon in the context of noncommutative field theory.

The NC space-time is characterized by,
\begin{equation}
[x^{\rho},x^{\sigma}]_{*}=i\theta^{\rho\sigma}.
\label{nc}
\end{equation}
The $*$-product is given by the Moyal-Weyl formula,
\begin{equation}
p(x)*q(x)=pq+\frac{i}{2}\theta^{\rho\sigma}\partial_{\rho}p\partial_{\sigma}q+~O(\theta^{2}).
\label{mw}
\end{equation}
All our discussions will be valid up to the first non-trivial order in $\theta$.

The NC extension of the Chern-Simons action has been derived in \cite{gs}.
The NC MCS model is defined in the following way \cite{of},
\begin{equation}
\hat {\cal {A}}^{\pm}_{MCS}=\int d^{3}x[-\frac{1}{4}\hat F^{\mu\nu}*\hat F_{\mu\nu}\pm \frac{M}{2}\epsilon^{\mu\nu\lambda}(\hat A_{\mu}*\partial_{\nu}\hat A_{\lambda}+\frac{2}{3}\hat A_{\mu}*\hat A_{\nu}*\hat A_{\lambda})],
\label{lmcs}
\end{equation}
where
$$\hat F_{\mu\nu}=\partial_{\mu}\hat A_{\nu}-\partial_{\nu}\hat A_{\mu}-i\hat A_{\mu}*\hat A_{\nu}+i\hat A_{\nu}*\hat A_{\mu}. $$
The structural similarity with corresponding expressions of
non-abelian gauge theory is very much apparent. It will be
revealed subsequently that the connection goes deeper.

Utilizing the Seiberg-Witten map, to the lowest non-trivial order
in $\theta$,
$$
\hat A_{\mu}=A_{\mu}+\theta^{\sigma\rho}A_{\rho}(\partial_{\sigma} A_{\mu}-\frac{1}{2}\partial_{\mu} A_{\sigma}),
$$
$$
\hat F_{\mu\nu}=F_{\mu\nu}+\theta^{\rho\sigma}(F_{\mu\rho}F_{\nu\sigma}-A_{\rho}\partial_{\sigma} F_{\mu\nu}),
$$
\begin{equation}
\hat\lambda =\lambda -\frac{1}{2}\theta^{\rho\sigma}A_{\rho}\partial_{\sigma}\lambda ,
\label{swm}
\end{equation}
(where $\hat\lambda$ and $\lambda$ are infinitesimal gauge
transformation parameters in NC and ordinary spacetimes), we
arrive at the following  $O(\theta)$ modified form of the NC MCS
theory, expressed in terms of ordinary spacetime variables
{\footnote {It should be kept in mind that in the conventional
Hamiltonian form of the field theory in (\ref{mcs}), only spatial
noncommutativity is allowed. But this is not of direct concern to
us in the present analysis.}} ,
$$
\hat {\cal {A}}^{\pm}_{MCS}=\int d^{3}x[-\frac{1}{4}(F^{\mu\nu}F_{\mu\nu}+2\theta^{\rho\sigma}(F^{\mu}_{~\rho}F^{\nu}_{~\sigma}F_{\mu\nu}
-\frac{1}{4}F_{\rho\sigma}F^{\mu\nu}F_{\mu\nu}))$$
\begin{equation}
\pm \frac{M}{2}\epsilon^{\mu\nu\lambda}A_{\mu}\partial_{\nu}A_{\lambda}],
\label{mcs}
\end{equation}
where $F_{\mu\nu}=\partial_{\mu}A_{\nu}-\partial_{\nu}A_{\mu}$. It
should be remembered that under the Seiberg-Witten map, the NC Chern-Simons
term exactly reduces to the Chern-Simons term in ordinary spacetime
\cite{gs}  to all orders in $\theta$. In 2+1-dimensions, (\ref{mcs}) further simplifies to,
$$
\hat {\cal {A}}^{\pm}_{MCS}=\int d^{3}x[-\frac{1}{4}F^{\mu\nu}F_{\mu\nu}\pm \frac{M}{2}\epsilon^{\mu\nu\lambda}A_{\mu}\partial_{\nu}A_{\lambda}$$
\begin{equation}
 -\frac{1}{8}\theta^{\rho\sigma}F_{\rho\sigma}F^{\mu\nu}F_{\mu\nu}].
\label{mmcs}
\end{equation}
The equations of motion are
\begin{equation}
\partial_{\mu}(F^{\mu\nu}\pm M\epsilon^{\mu\nu\lambda}A_{\lambda})=O(\theta),
\label{eqm}
\end{equation}
where explicit form of the $O(\theta)$ term in the right hand side
is not required for our present analysis. The change in $A_{\mu}$
\begin{equation}
\delta A_{\mu}=a_{\mu},
\label{tr}
\end{equation}
induces the following changes in the actions,
\begin{equation}
\delta \hat {\cal {A}}^{\pm}_{MCS}=\int d^{3}x[(J^{\pm}_{\mu\nu}+J^{(\theta)}_{\mu\nu})\partial^{\mu}a^{\nu}],
\label{cur}
\end{equation}
where,
\begin{equation}
J^{\pm}_{\mu\nu}\equiv -F_{\mu\nu}\pm M\epsilon^{\mu\nu\lambda}A_{\lambda},
\label{jpm}
\end{equation}
\begin{equation}
J^{(\theta)}_{\mu\nu}\equiv -\frac{1}{4}(F^{2}\theta_{\mu\nu}+2(\theta .F)F_{\mu\nu}).
\label{jnc}
\end{equation}
A short hand notation $\theta^{\mu\nu}F_{\mu\nu}=\theta .F$ has been adopted.

Now the auxiliary variable $C_{\mu\nu}$ is introduced in the
action via the background interaction and contact terms,
\begin{equation}
\tilde {\cal {A}}=\hat{\cal {A}} -\int \frac{1}{2}[(J+J^{(\theta)}).C+\frac{1}{2}C^{2}].
\label{lb}
\end{equation}
The transformation  {\footnote{We have checked that keeping $\delta C_{\mu\nu}=\partial_{\mu}a_{\nu}-\partial_{\nu}a_{\mu}$ and introducing more interference terms in the action does not solve the problem at hand.}} of $C_{\mu\nu}$,
\begin{equation}
\delta C_{\mu\nu}=\partial_{\mu}a_{\nu}-\partial_{\nu}a_{\mu}-\delta J^{(\theta)}_{\mu\nu},
\label{delb}
\end{equation}
together with the transformation of $A_{\mu}$ in (\ref{tr}), changes the actions by,
\begin{equation}
\delta \hat {\cal {A}}^{\pm}_{MCS}=\int d^{3}x [\pm \frac{M}{2}C^{\mu\nu}\epsilon_{\mu\nu\lambda}a^{\lambda}+\frac{1}{2}(-F^{\mu\nu}\pm M\epsilon^{\mu\nu\lambda}A_{\lambda})\delta J^{(\theta)}_{\mu\nu}].
\label{dela}
\end{equation}
The idea behind introduction of the fine tuned
$C_{\mu\nu}$-dependent counterterms in the two actions (that are to be
soldered) is that the variations in the two actions  will be of
opposite signature. This feature will give rise to the invariance
in the soldered action. In the present instance, this is true for
$\theta_{\mu\nu}=0$ but this does not quite happen for non-zero
$\theta_{\mu\nu}$ since the $F_{\mu\nu}$-term appears with the
same sign in variations of both the actions. However, notice that
considerations of the equations of motion (\ref{eqm}) reveals that
the factor $(-F^{\mu\nu}\pm M\epsilon^{\mu\nu\lambda}A_{\lambda})$
itself in (\ref{dela})  is of $O(\theta)$ and so
$\frac{1}{2}(-F^{\mu\nu}\pm
M\epsilon^{\mu\nu\lambda}A_{\lambda})\delta J^{(\theta)}_{\mu\nu}$
in (\ref{dela}) will contribute to $O(\theta^{2})$ correction.
Thus, this term in the variations of the actions in (\ref{dela})
can be ignored in our present discussion. The rest of the
variations in (\ref{dela}) are of the form that is amenable to
soldering. In fact the variations are same (up to $O(\theta))$ as
their ordinary spacetime counterpart.

Let us pause for a moment to appreciate the  closeness of our
analysis with existing results in the context of soldering of
non-abelian self dual models \cite{iw}. First of all, the Euler
kernel, that is the Noether current receives non-linear
contributions as a result of noncommutativity (or non-abelian
nature \cite{iw}), which is quite natural. However, it is striking
that the auxiliary field $C_{\mu\nu}$ ceases to transform in the
conventional way and the more involved transformation (\ref{delb})
is clearly "identical" to its counterpart in the non-abelian
context \cite{iw}.

We introduce the auxiliary $C_{\mu\nu}$ field and write down the total action as,
$$
\hat {\cal{A}}=\tilde {\cal{A}}^{+}(A,C)+\tilde {\cal{A}}^{-}(B,C)$$
\begin{equation}
=\hat{\cal{A}}^{+}(A)+\hat{\cal{A}}^{-}(B)
-\frac{1}{2}\int[C^{2}+\{J^+(A)+J^-(B)+J^{(\theta)}(A)+J^{(\theta)}(B)\}.C].
\label{ab}
\end{equation}
The $C_{\mu\nu}$ field is constrained by the relation,
\begin{equation}
C_{\mu\nu}=-\frac{1}{2}(J_{\mu\nu}^+(A)+J_{\mu\nu}^-(B)+J_{\mu\nu}^{(\theta)}(A)+J_{\mu\nu}^{(\theta)}(B)).
\label{b}
\end{equation}
This allows us to replace $B_{\mu\nu}$ in favor of the basic
dynamical variables and we obtain the soldered Lagrangian,
$$
\hat {\cal {L}}^{(S)}=-\frac{1}{8}F^{\mu\nu}(A-B)F_{\mu\nu}(A-B)+\frac{M^{2}}{4}(A-B)^{\mu}(A-B)_{\mu}
$$
$$
-\frac{1}{16}[(-F(A)^{\mu\nu}+M\epsilon^{\mu\nu\lambda} A_{\lambda})+(-F^{\mu\nu}(B)-M\epsilon^{\mu\nu\lambda} B_{\lambda}]$$
\begin{equation}
[F^2(A)\theta_{\mu\nu} +2(\theta .F(A))F_{\mu\nu}(A)+F^2(B)\theta_{\mu\nu} +2(\theta .F(B))F_{\mu\nu}(B)].
\label{llsm}
\end{equation}
It should be mentioned that the above step is somewhat formal
since the theory is obviously not Gaussion. However, this is
legitimate as far as classical considerations go. Similar steps
have also been performed in \cite{iw}.

Note that the $\theta$-independent part of the action in
(\ref{llsm}) is in the form that was advertised at the beginning.
In fact, this part is identical to the ordinary spacetime result
\cite{bw}. The remaining part of the action $\hat L^{(S)}$ in
(\ref{llsm}) is once again dropped since it is of $O(\theta^{2})$.
The argument is exactly the same as the one given below
(\ref{dela}). Hence the soldered theory is the Proca term with
$O(\theta^{2})$ modification,
$$
\hat{\cal{ L}}^{(S)}=-\frac{1}{8}F^{\mu\nu}(A-B)F_{\mu\nu}(A-B)+\frac{M^{2}}{4}(A-B)^{\mu}(A-B)_{\mu} +O(\theta^{2})
$$
\begin{equation}
=-\frac{1}{8}F^{\mu\nu}(G)F_{\mu\nu}(G)+\frac{M^{2}}{4}G^{\mu}G_{\mu} +O(\theta^{2}),
\label{11}
\end{equation}
where, in keeping with our earlier notation in (\ref{s4}),
$G\equiv A-B$. Thus our major observation  is that, to the lowest
non-trivial order in $\theta$,  soldering of the NC MCS models is
indeed possible and consistent. The whole process is clearly
reminiscent of the similar phenomenon \cite{bw} in ordinary
spacetime, with the noncommutativity introducing a non-abelian
flavor.

We are now faced with the question that how far is it justified to
identify the above massive vector model, whose kinetic part is
{\it{identical to the ordinary spacetime Maxwell term}}, and not
of the NC Maxwell form, as the NC generalization of the Proca
model, up to $O(\theta)$. Our views, favoring the above
identification are presented below.

There are some points in the present work that need to be stressed
in the perspective of the recent paper \cite{bot}. First of all,
in the present work, we have taken the noncommutative
generalization of  MCS theory that has been suggested in \cite{of}
and advocated in \cite{bot}. To $O(\theta)$, we recover the action
that is same as that of the ordinary spacetime Proca theory. We
claim this to be the {\it{noncommutative generalization of the
Proca theory}} as well. Our contention is that since there is no
gauge invariance in the Proca model which is a free theory having
only quadratic terms in the action, there will be no effects of
noncommutativity, at least to $O(\theta)$, assuming that total
derivative terms in the action need not be taken in to account. This ties up
very nicely with our previous work \cite{sg1}, where similar
reasonings were put forward for the noncommutative extension of
the self-dual model. This idea is firmly based on the basic
premises of the Seiberg-Witten map where non-trivial nature of the
mapping comes in to play {\it{only in the presence of gauge
invariance}}.

This means that, to $O(\theta )$, the model in (\ref{11}) can be
elevated to the action of the corresponding NC model,
\begin{equation}
\hat{\cal{ S}}^{(S)}=\int d^3x \hat{\cal{ L}}^{(S)}=\int d^3x (
-\frac{1}{8}\hat F^{\mu\nu}(\hat G)*\hat F_{\mu\nu}(\hat
G)+\frac{M^{2}}{4}\hat G^{\mu}*\hat G_{\mu}), \label{ss}
\end{equation}
with the identification $\hat G=G,~\hat F^{\mu\nu}(\hat G)\equiv
\partial ^\mu \hat G^\nu - \partial ^\nu \hat G^\mu $, since there is
no non-trivial Seiberg-Witten map for non-gauge theories. Thus
$\theta$-dependent terms in (\ref{ss}) can only appear from the
$*$-products. However, the theory being quadratic, the $O(\theta
)$ contributions coming from the $*$-products are total
derivatives and will vanish in the action and we are left with,
$$
\hat{\cal{ S}}^{(S)}=\int d^3x ( -\frac{1}{8}\hat F^{\mu\nu}(\hat
G)\hat F_{\mu\nu}(\hat G)+\frac{M^{2}}{4}\hat G^{\mu}\hat G_{\mu})
+O(\theta ^2)$$
\begin{equation}
=\int d^3x ( -\frac{1}{8} F^{\mu\nu}(G)
F_{\mu\nu}(G)+\frac{M^{2}}{4} G^{\mu} G_{\mu}) +O(\theta ^2) ,
\label{ss1}
\end{equation}
where in the last step we have recovered (\ref{11}). Hence,
(\ref{ss}) (or effectively (\ref{11})) is the cherished form of
the noncommutative Self Dual model.

Furthermore, from the soldering point of view, the above criterion
corroborates with the ordinary spacetime result where self-dual
and anti self-dual models can also be soldered to generate the
Proca model. This has to be the case since the self-dual models
are dual to the MCS models respectively. In the present paper, we
have demonstrated the noncommutative counterpart of MCS-Proca
soldering. On the other hand we can consider the noncommutative
generalization of the  soldering of NC SD models to generate NC
Proca model. But, as we have argued, to $O(\theta)$ there will be
no changes in {\it{any}} of the models participating in the
soldering process, since all of them are quadratic theories and
none of them are gauge theories. Hence we will reach the result,
identical to the ordinary spacetime one, which also agrees with
the conclusion presented here. This demonstrates the robustness of
the soldering programme as well as consistency in our way of
defining noncommutative generalizations of self-dual or Proca
theories.

Indeed it would be interesting if the noncommutative soldering can
be performed consistently for the parent and daughter models where
{\it{all}} of them are gauge theories. Such a problem has been
discussed in \cite{bw} in the context of electromagnetic duality.
Its NC extension is under study. \vskip 1cm
{\it{Acknowledgement}}: I thank the Referee for raising pertinent
questions which have led to an overall improvement of the paper.

\end{document}